\documentclass[journal]{IEEEtran}

\usepackage{graphicx, subfigure}
\usepackage{amsmath,amssymb,amsfonts}
\usepackage{algorithmic}
\usepackage{algorithm}
\usepackage{caption}
\usepackage{hyperref}
\usepackage[capitalise]{cleveref}
\usepackage{float}
\usepackage{enumerate}
\usepackage{amsmath}
\usepackage{cite}
\usepackage{bm}

\def\BibTeX{{\rm B\kern-.05em{\sc i\kern-.025em b}\kern-.08em
    T\kern-.1667em\lower.7ex\hbox{E}\kern-.125emX}}

\usepackage{acro}[=v3]
\acsetup{
    first-style = long-short,
    list/display = used,
    pages/display = first
}

\DeclareAcronym{SPRP}{
    short=SPRP,
    long= semantic-based preemptive resource provisioning
}

\DeclareAcronym{SRCM}{
    short=SRCM,
    long= semantic-based resource computation \& mapping
}

\DeclareAcronym{SSCI}{
    short=SSCI,
    long= semantic-based service class identification
}

\DeclareAcronym{SPSS}{
    short=SPSS,
    long= semantic-based proactive slice switching
}
\DeclareAcronym{MILP}{
    short=MILP,
    long= mixed integer linear programming
}
\DeclareAcronym{3GPP}{
    short=3GPP,
    long=3rd generation partnership project
}

\DeclareAcronym{UE}{
    short=UE,
    long=user equipment
}
\DeclareAcronym{SCM}{
  short=SCM,
  long=structural causal model 
}
\DeclareAcronym{D2D}{
    short=D2D,
    long=device-to-device
}

\DeclareAcronym{NMSE}{
  short=NMSE,
  long=normalized mean square error
}
\DeclareAcronym{GPS}{
    short=GPS,
    long=global positioning system
}

\DeclareAcronym{PTT}{
  short=PTT,
  long=push-to-talk 
}

\DeclareAcronym{CCTV}{
    short=CCTV,
    long=closed-circuit television
}

\DeclareAcronym{GSMA}{
 short=GSMA,
 long=global system for mobile communications
}
\DeclareAcronym{CLA}{
 short=CLA,
 long=closed loop automation
}
\DeclareAcronym{AVG}{
 short=AVG,
 long=average 
}
\DeclareAcronym{FL}{
 short=FL,
 long=federated learning
}

\DeclareAcronym{FDD}{
    short=FDD,
    long=Frequency Division Duplex
}

\DeclareAcronym{SMO}{
    short=SMO,
    long=Service Management and Orchestration
}

\DeclareAcronym{DNS}{
    short=DNS,
    long=dynamic network slicing
}

\DeclareAcronym{VNF}{
    short=VNF,
    long=virtual network function
}

\DeclareAcronym{NIC}{
    short=NIC,
    long=Network Interface Card
}

\DeclareAcronym{NF}{
    short=NF,
    long= Network Function
}

\DeclareAcronym{NFV}{
    short=NFV,
    long=network function virtualization
}

\DeclareAcronym{SBA}{
    short=SBA,
    long=service based architecture
}

\DeclareAcronym{MVNO}{
    short=MVNO,
    long=mobile virtual network operator
}

\DeclareAcronym{AIMLFW}{
    short=AIMLFW,
    long=\ac{AI}\ac{ML} Framework
}

\DeclareAcronym{DSNS}{
    short=DSNS,
    long=Dynamic Semantic-Based Network Slicing
}
\DeclareAcronym{FCAI}{
    short=FCAI,
    long=Finnish Center for Artificial Intelligence
}

\DeclareAcronym{QoS}{
    short=QoS,
    long=quality-of-service
}

\DeclareAcronym{AoF}{
    short=AoF,
    long=Academy of Finland
}
\DeclareAcronym{QoE}{
    short=QoE,
    long=quality-of-experience 
}

\DeclareAcronym{UPF}{
    short=UPF,
    long=user plane function
}

\DeclareAcronym{NSSF}{
    short=NSSF,
    long=network slice selection function
}

\DeclareAcronym{AMF}{
    short=AMF,
    long=access and mobility management function
}

\DeclareAcronym{NRF}{
    short=NRF,
    long=network repository function 
}
\DeclareAcronym{PCF}{
    short=PCF,
    long=policy control function 
}
\DeclareAcronym{AUSF}{
    short=AUSF,
    long=authentication server function 
}
\DeclareAcronym{UDM}{
    short=UDM,
    long=unified data management 
}

\DeclareAcronym{SMF}{
    short=SMF,
    long=session management function
}

\DeclareAcronym{S-NSSAI}{
    short=S-NSSAI,
    long=single network slice selection assistance information
}

\DeclareAcronym{NSSAI}{
    short=NSSAI,
    long=network slice selection assistance information
}

\DeclareAcronym{SST}{
    short=SST,
    long=slice/service type
}
  \DeclareAcronym{SD}{
    short=SD,
    long=slice differentiator
}

\DeclareAcronym{XR}{
    short=XR,
    long= extended reality 
}
\DeclareAcronym{VR}{
    short=VR,
    long= Virtual Reality 
}
\DeclareAcronym{AR}{
    short=AR,
    long= Augmented Reality 
}
\DeclareAcronym{MR}{
    short=MR,
    long= Mixed Reality 
}

\DeclareAcronym{MEC}{
    short=MEC,
    long=Mobile Edge Computing
}

\DeclareAcronym{RAT}{
    short=RAT,
    long=radio access technology 
}

\DeclareAcronym{TDD}{
    short=TDD,
    long=Time Division Duplex
}

\DeclareAcronym{VTRC}{
    short=VTRC,
    long=Virginia Tech Research Center
}
\DeclareAcronym{PAWR}{
    short=PAWR,
    long=Platforms for Advanced Wireless Research  
}
 \DeclareAcronym{NSF}{
    short=NSF,
    long=National Science Foundation 
}
\DeclareAcronym{VT}{
    short=VT,
    long=Virginia Tech
}
\DeclareAcronym{UOULU}{
    short=UOulu,
    long=University of Oulu
}
\DeclareAcronym{SEMA}{
 short=SEMA,
 long=state emergency management agency
}
\DeclareAcronym{RRC}{
 short=RRC,
 long=radio resource control
}
\DeclareAcronym{NAS}{
 short=NAS,
 long=non-access stratum
}
\DeclareAcronym{ReD}{
    short=R\&D,
    long=Research and Development
}
\DeclareAcronym{NS}{
    short=NS,
    long=network slicing
}
\DeclareAcronym{REU}{
    short=REU,
    long=Research Experiences for Undergraduates
}

\DeclareAcronym{TAMUS}{
   short=TAMUS,
   long=Texas A\&M University System
}

\DeclareAcronym{NEARRTRIC}{
    short=Near-RT RIC,
    long=Near Real Time RIC
}

\DeclareAcronym{NONRTRIC}{
    short=Non-RT RIC,
    long=Non Real Time RIC
}

\DeclareAcronym{CSP}{
    short=CSP,
    long=Communication Service Providers
}

\DeclareAcronym{OPENRAN}{
    short=Open RAN,
    long=Open Radio Access Network
}

\DeclareAcronym{UoO}{
    short=UOulu,
    long=University of Oulu
}

\DeclareAcronym{EUTRAN}{
    short=E-UTRAN,
    long=Evolved Universal Terrestrial RAN
}

\DeclareAcronym{CCI}{
    short=CCI,
    long=Commonwealth Cyber Initiative
}

\DeclareAcronym{GDPR}{
    short=GDPR,
    long=General Data Protection Regulation
}

\DeclareAcronym{EU}{
    short=EU,
    long=European Union
}
\DeclareAcronym{HAR}{
 short=HAR,
 long=human activity recognition
}
\DeclareAcronym{ECE}{
 short=ECE,
 long= Electrical and Computer Engineering
}
\DeclareAcronym{E2E}{
    short=E2E,
    long=end-to-end
}
\DeclareAcronym{LLM}{
 short=LLM,
 long=large language models
}

\DeclareAcronym{uRLLC}{
    short=uRLLC,
    long=ultra reliable and low latency communication
}

\DeclareAcronym{mMTC}{
    short=mMTC,
    long=massive machine type communication
}

\DeclareAcronym{eMBB}{
    short=eMBB,
    long=enhanced mobile broadband
}

\DeclareAcronym{RIC}{
    short=RIC,
    long=Radio Intelligent Controller
}

\DeclareAcronym{MNO}{
    short=MNO,
    long=Mobile Network Operator
}

\DeclareAcronym{UTRAN}{
    short=UTRAN,
    long=Universal Radio Access Network
}

\DeclareAcronym{VRAN}{
    short=VRAN,
    long=Virtualized Radio Access Network
}

\DeclareAcronym{ORAN}{
    short=O-RAN,
    long=Open Radio Access Network
}

\DeclareAcronym{ORANSC}{
    short=O-RAN SC,
    long=O-RAN Software Community
}

\DeclareAcronym{ONAP}{
    short=ONAP,
    long=Open Networking Automation Platform
}

\DeclareAcronym{SDN}{
    short=SDN,
    long=software defined networking 
}

\DeclareAcronym{SDR}{
    short=SDR,
    long=Software Defined Radio 
}

\DeclareAcronym{RAN}{
    short=RAN,
    long=radio access network
}

\DeclareAcronym{FCC}{
    short=FCC,
    long=Federal Communication Commission
}

\DeclareAcronym{LTE}{
    short=LTE,
    long=Long-Term Evolution
}
\DeclareAcronym{SLA}{
    short=SLA,
    long=service level agreement
}
\DeclareAcronym{4G}{
    short=4G,
    long=Fourth Generation
}
\DeclareAcronym{CN}{
    short=CN,
    long=core network
}
\DeclareAcronym{TN}{
    short=TN,
    long=transport network
}
\DeclareAcronym{5G}{
    short=5G,
    long=Fifth Generation
}
\DeclareAcronym{HD}{
    short=HD,
    long=High Definition
}

\DeclareAcronym{FPS}{
    short=FPS,
    long=Frame Per Second
}

\DeclareAcronym{6G}{
    short=6G,
    long=Sixth Generation
}

\DeclareAcronym{NR}{
    short=NR,
    long=New Radio
}
\DeclareAcronym{5GTN}{
    short=5GTN,
    long=5G Test Network
}

\DeclareAcronym{nonrtric}{
    short=Non-RT RIC,
    long= Non Real-Time Radio Intelligent Controller
}

\DeclareAcronym{neartric}{
    short=Near-RT RIC,
    long= Near Real-Time Radio Intelligent Controller
}

\DeclareAcronym{AI}{
    short=AI,
    long=  artificial intelligence
}

\DeclareAcronym{ML}{
    short=ML,
    long=  machine learning
}
\DeclareAcronym{MLOps}{
    short=MLOps,
    long=  Machine Learning Operations
}

\DeclareAcronym{RL}{
    short=RL,
    long=  reinforcement learning
}

\DeclareAcronym{OAI}{
    short=OAI,
    long=OpenAirInterface
}

\DeclareAcronym{WLAN}{
    short=WLAN,
    long=Wireless Local Area Network
}

\DeclareAcronym{SRS}{
    short=SRS,
    long=Software Radio System
}
\DeclareAcronym{FoV}{
    short=FoV,
    long=Field of View 
}

\DeclareAcronym{PMP}{
    short=PMP,
    long=Project Management Professional 
}
\DeclareAcronym{PMI}{
    short=PMI,
    long=Project Management Institute 
}
\DeclareAcronym{PMBOK}{
    short=PMBOK,
    long=Project Management Body of Knowledge 
}

\DeclareAcronym{OSP}{
    short=OSP,
    long=Office of Sponsored Programs
}

\DeclareAcronym{GRA}{
    short=GRA,
    long=Graduate Research Assistant
}

\DeclareAcronym{API}{
    short=API,
    long=application programming interface
}
\DeclareAcronym{JSC}{
  short=JSC,
  long=joint source and channel coding
}
\DeclareAcronym{NSI}{
    short=NSI,
    long=network slice instance
}
\DeclareAcronym{DL}{
    short=DL,
    long=deep learning
}
\DeclareAcronym{KPI}{
    short=KPI,
    long=key performance indicator
}

\DeclareAcronym{SRA}{
    short=SRA,
    long=strategic research area
}

\DeclareAcronym{SRA3}{
    short=SRA3,
    long=strategic research area 3
}

\DeclareAcronym{MANO}{
    short=MANO,
    long=Management and Network Orchestration
}

\DeclareAcronym{CPRA}{
    short=CPRA,
    long=California Privacy Rights Act 
}

\DeclareAcronym{IoT}{
    short=IoT,
    long=Internet of Things
}
\DeclareAcronym{OTIC}{
 short=OTIC,
 long=Open Testing Integration Center
}

\DeclareAcronym{CSMF}{
    short=CSMF,
    long=communication service management function
}

\DeclareAcronym{NSMF}{
    short=NSMF,
    long= network slice management function
}

\DeclareAcronym{NSSMF}{
    short=NSSMF,
    long= network slice subnet management function
}

\DeclareAcronym{CSC}{
    short=CSC,
    long= communication service customer
}

\DeclareAcronym{NSP}{
    short=NSP,
    long= network slice provider
}

\DeclareAcronym{NSA}{
    short=NSA,
    long= network slice agent
}

\DeclareAcronym{NST}{
    short=NST,
    long= network slice template
}

\begin{document}
\title{A Framework for AI-Native Semantic-Based Dynamic Slicing for 6G Networks}
%




\author{
\IEEEauthorblockN{Mayukh Roy Chowdhury\IEEEauthorrefmark{5},
Eman Hammad\IEEEauthorrefmark{2},
Lauri Loven\IEEEauthorrefmark{3}, \\
Susanna Pirttikangas\IEEEauthorrefmark{4},
Aloizio P da Silva\IEEEauthorrefmark{1}, and
Walid Saad \IEEEauthorrefmark{1}
}

\IEEEauthorblockA{
\IEEEauthorrefmark{1} Virginia Tech, Arlington, VA, USA\
Email: {aloiziops, walids}@vt.edu \\
\IEEEauthorrefmark{2}Texas A\&M University \
Email: {eman.hammad}@tamu.edu \\
\IEEEauthorrefmark{3}Center for Ubiquitous Computing, University of Oulu, Finland\
Email: {lauri.loven}@oulu.fi \\
\IEEEauthorrefmark{4}AMD Silo AI, Finland\
Email: susanna.pirttikangas@amd.com \\
\IEEEauthorrefmark{5}Nokia Bell Labs\
Email: mayukh.roy\_chowdhury@nokia-bell-labs.com 
}}
\maketitle
\begin{abstract}
In the ensuing ultra-dense and diverse environment in future \ac{6G} communication networks, it will be critical to optimize network resources via mechanisms that recognize and cater to the diversity, density, and dynamicity of system changes. However, coping with such environments cannot be done through the current network approach of compartmentalizing data as distinct from network operations. Instead, we envision a computing continuum where the content of the transmitted data is considered as an essential element in the transmission of that data, with data sources and streams analyzed and distilled to their essential elements, based on their semantic context, and then processed and transmitted over dedicated slices of network resources. By exploiting the rich content and semantics within data for dynamic and autonomous optimization of the computing continuum, this article opens the door to integrating communication, computing, cyber-physical systems, data flow, and AI, presenting new and exciting opportunities for cross-layer design. We propose semantic slicing, a two-pronged approach that builds multiple virtual divisions within a single physical and data infrastructure, each with its own distinct characteristics and needs. We view semantic slicing as a novel shift from current static slicing techniques, extending existing slicing approaches such that it can be applied dynamically at different levels and categories of resources in the computing continuum. Further it propels the advancement of semantic communication via the proposed architectural framework. 
\end{abstract}



\section{Introduction}

Upcoming 6G communication networks are anticipated to extend 5G usage scenarios with ubiquitous connectivity, AI and communication, and integrated sensing and communication, serving a massive number of both human and machine-type users within an ever more diverse and dense communication and sensing infrastructure. Consequently, \ac{6G} wireless networks are envisioned to add a wide spectrum of new service classes and, along with them, diverse and dynamic \ac{QoS} requirements \cite{6GVision_NetMag20}. 
Those \ac{QoS} requirements often translate to \ac{QoE} requirements that should capture the actual experience of the users. Example \ac{QoE} dimensions include ease of access to the service, time taken for service execution, comprehensibility of the system's functioning, the quality of sensor data or media, and the effectiveness of the system's decision-making processes \cite{Alqerm2021, Lu2016}. 

Operators are able to support some of the foreseen 6G \ac{QoS} (and by extension QoE) needs, using an approach called \Ac{NS}. With \ac{NS}, an operator can provide customized services through the creation of multiple isolated logical networks in a shared physical network antamud resources \cite{NS5G_WCommMag19}. This creates great benefits in terms of efficient resource utilization, customized service guarantees, and possibly higher levels of security through isolation. Current \ac{NS} state-of-the-art approaches rely on a flow that starts by a high-level categorization of \ac{QoS} requirements (i.e. latency, throughput, packet loss), and then network resources are orchestrated to support this high-level categorization which remains constant throughout. Hence, this proved to hinder the efficient utilization of resources and to significantly abstract other critical requirements. For example, merely looking at whether the minimum throughput requirement is satisfied or not, may not be sufficient; instead, it might be more useful to quantify how well the meaning of the transmitted message is conveyed to the receiver while spending the minimum possible resources.

Considering the \emph{semantics} of the captured and transmitted data can bring benefits not only to users but also to the systems supporting the \ac{QoS} delivery. As defined in \cite{lessdata_walid_arxiv22}, the semantics of data are primarily ``structures'' that can be discovered inside a dataset, to represent the overall ``meaning'' of information in a minimalist way. Those structures, often called semantic content elements (shorthanded ``semantics'' hereinafter), essentially summarize the entire content of the data of interest. 

Supporting diverse \ac{QoS} requirements through different slices is a complex task, particularly if those requirements change over time. To this end, there is an opportunity to leverage advanced methods such as \ac{AI} to analyze the data and extract useful semantics that can be used for allocating network slice resources in a dynamic manner. Exploiting data semantics in communication is not a new concept, and has been an active area of research which, traditionally, focused primarily on more intelligent approaches to ``data compression''. \textit{Semantic slicing} is a new approach for exploiting semantics to capture and support the dynamically changing and diverse \ac{QoS}. There are currently no means to perform such semantic slicing, a technique that would allow for the separation of concepts encoded within the data generated in the edge and use them for intelligent resource management across slices.

Semantic slicing introduces a fundamental shift in how communications systems interact with and support the data they deliver and the applications they serve. Several possible applications can be realized based on semantic slicing including more efficient and effective management of data and resources, as well as a much finer granularity of control on access to data and network services.

The main contribution of this paper is a holistic architecture for implementing a new semantic slicing framework. Compared to the existing \ac{NS} technologies, the proposed semantic slicing framework brings in the following benefits:

\begin{enumerate}
     
   \item {\textbf{Outlining a practical approach for implementing Semantic communication:}} Semantic slicing provides a framework that supports semantic communication, i.e. allocates network resources for sending not the whole data, but a semantic representation of it. It also paves the way for a new range of \ac{QoS} parameters to be supported by the slice providers as part of the \ac{SLA}, which is expected to enhance the user experience like never before. It goes beyond the application or context-aware slicing as it involves finding structures or representations from the data that goes past the application boundaries.

   \item {\textbf{Enhancing network management through a reasoning plane:}} \ac{5G} networks inherently support separation between control and data planes. Semantic slicing can prove to be a way forward in that path by bringing in a reasoning plane (see \cite[Figure 12]{lessdata_walid_arxiv22} between the data and the control plane in \ac{6G} networks. By analyzing the data, semantic representation can be extracted which can govern the policies that the control plane will take. Semantics will also help reason any failure in the network or why a slice is not able to maintain a certain level of \ac{QoS}, so that the slice orchestrator can take appropriate actions autonomously.
   
   \item {\textbf{Adopting \ac{AI}-driven closed-loop automation of slice orchestration:}} Given the range of diverse applications and heterogeneous \ac{QoS} classes \ac{6G} is expected to support, there is a critical need to minimize manual intervention and enable closed-loop automation in network slice orchestration. Existing works in the area of \ac{DNS} discuss scaling up or down network resources dynamically, but the policies governing those decisions are mostly static and require (manual) human intervention. The proposed semantic slicing architecture will utilize \ac{AI}-powered close-loop feedback capabilities to automate the provisioning and management of network resources leveraging semantics.
      
\end{enumerate}

The remainder of this article is organized as follows: \cref{sec:background} provides a brief background of semantic slicing. \cref{sec:motivation,sec:related} describe some motivating scenarios and existing works in this direction, respectively. 
In  \cref{sec:5gslicing}, we outline the \ac{3GPP} suggested architecture for network slicing in service-based architecture of 5G.
The additional components required for supporting the proposed semantic slicing are introduced in \cref{sec:semantic} while a blueprint for the same is explained using a realistic use case in \cref{sec:usecase}. Subsequently, in Section \ref{sec:init_results},  some initial results in this direction are presented. Finally, we discuss some of the open challenges and research problems in this domain in \cref{sec:challenges}, followed by a conclusion in \cref{sec:concl}.


\section{Background}\label{sec:background}

The concept of transmitting the ``meaning'' or semantics of messages dates back to the work of Weaver who suggested that incorporating the meaning of messages and their effect on a system during the communication between a transmitter and a receiver could potentially lead to better communication performance. However, that concept did not materialize at the time of Weaver. Recently, the interest in semantic communication was reinvigorated, in part due to recent advances in \ac{AI}. The majority of the prior art in this space, currently defines semantic communications as a form of \ac{JSC}, whose input are features extracted from the source data through the use of \ac{DL} techniques. Under this view, semantic communication essentially boils down to a form of \ac{DL}-assisted compression or \ac{JSC}. However, as pointed out in \cite{lessdata_walid_arxiv22}, while important, this conventional view may not realize the full potential of semantic communications. Instead, one can take an \ac{AI}-native perspective on the problem, and view semantic communication as a means to enable ``human-like'' communication between transmitter and receiver. Here, the transmitter extracts data structures, called semantic content elements, that summarize the entire content of a data set and, then, uses those structures to create a semantic language between the transmitter and receiver, through techniques like causal reasoning. Those structures must be minimalist, generalizable (i.e., they can be used to describe different context and information), and efficient. In this view, the communicating nodes can now learn to acquire an underlying understanding (i.e., a model) of the information being communicated and, then, use this understanding to not only minimize data transfer, but also improve reliability, interoperability, and context awareness. This understanding of the underlying information allows semantic communication systems to exploit ``compute'' resources to compensate for communication resources when the latter are unavailable.

\begin{figure}
	\centering
   \includegraphics[width=\columnwidth]{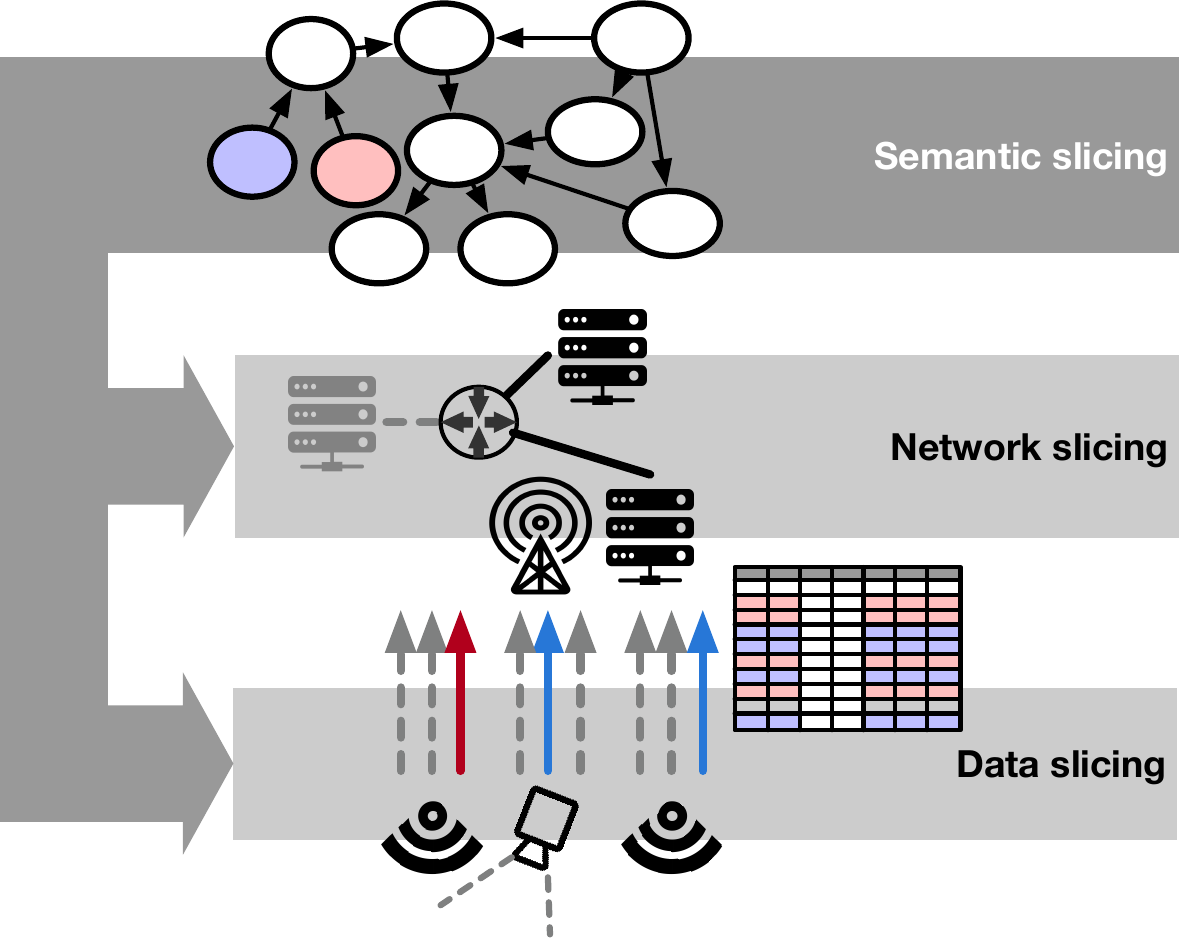}
    \caption{Two functions of semantic slicing: data slicing and network slicing.}
  \label{fig:example}
\end{figure}

Semantic slicing can fall under the umbrella of the second view on semantic communication above. In this context, semantic slicing combines two major functions: (i) network slicing, which provisions compute and communication resources, and (ii) data slicing, which dynamically partitions and manages data resources in the compute continuum (\cref{fig:example}). Traditionally, data slicing incorporates methods to extract subsets of data from larger datasets based on specific conditions or criteria. 
Data analysis and pre-processing leverage different data slicing techniques to highlight relevant information and decrease data complexity for programming and databases, for more efficient data analysis and enhanced privacy, among other benefits.
Nonetheless, existing data slicing methods do not consider data architectures in the compute continuum, and the slicing of the data streams and processing within that continuum. 
In the context of semantic slicing, we use data slicing to  
partition data into smaller subsets in the data pipelines between the nodes in the compute continuum (e.g., \ac{IoT} gateways, fog nodes, edge servers) as well as the data processing chains within those nodes. The data partitioning can be based on different criteria such as time, the order or sequence of an observation, raw features, or predicted features. Each subset can then be transmitted and further processed over a particular network slice with provisioned resources. 

Resource provisioning in conventional \ac{NS} is relatively static, often leading to either slice overloading or resource under-utilization. \ac{DNS} offers methods for the dynamic management of compute and communication resources based on slice demand. As future networks will be highly heterogeneous and dynamic, maintaining \ac{SLA} adherence requires continuous monitoring of resources and their real-time orchestration, informed by user application data analytics and feedback. Moreover, slice users such as \acp{MVNO}, which may need to refine coarse-grained \acp{NS} into finer-grained ones to adhere to the requirements of their clients, further increase the need for dynamic, \ac{AI} powered orchestration of \ac{DNS}.

The basis of semantic slicing lies in the \textit{semantic context}, that is, the state of the semantic elements of the application as well as the resources used to process and transmit those elements within the application context. In consequence, semantic slicing comprises the real-time, intelligent allocation and release of network resources, aligned with traffic patterns, user priorities, and \ac{QoS} stipulations of \ac{SLA}, systematically partitioning the components in the compute continuum: data, edge compute, storage, \ac{RAN}, \ac{CN}, and \ac{TN}.

To summarize, the objective of semantic slicing is to provide enhanced \ac{QoS} and \ac{QoE} by optimizing \ac{DNS} based on the semantic context. In the envisioned architecture, a semantic slice thrives for the following: (i) more fine-tuned resource allocation and its mapping to QoS, (ii) more dynamic and adaptive response to system-level and application/data level changes, (iii) allows the infrastructure to cater for new and different QoS metrics (in the task mapping).
It also manages the optimization of the whole \ac{E2E} data pipeline, personalizing \ac{E2E} data ownership and creating customized user services based on the content of the data as well as application requirements. This allows for efficient, effective and more controlled data processing and analysis, as well as more effective management of network resources and sharing of data and consent. 


\section{Motivating scenarios}\label{sec:motivation}

New \ac{6G} applications like the metaverse, \ac{XR}, telehealth, and smart transportation, each with distinct \ac{QoS} requirements and unique "semantic" structures, are prime candidates for semantic slicing. They can be categorized into four main use cases: \textit{prioritizing data transfers} (e.g., first responder), \textit{segregating data transmissions} (e.g., financial services), \textit{managing resources} (e.g., telehealth), and \textit{managing consent} (e.g., telepresence).

\begin{figure*}
	\centering
   \includegraphics[width=\textwidth]{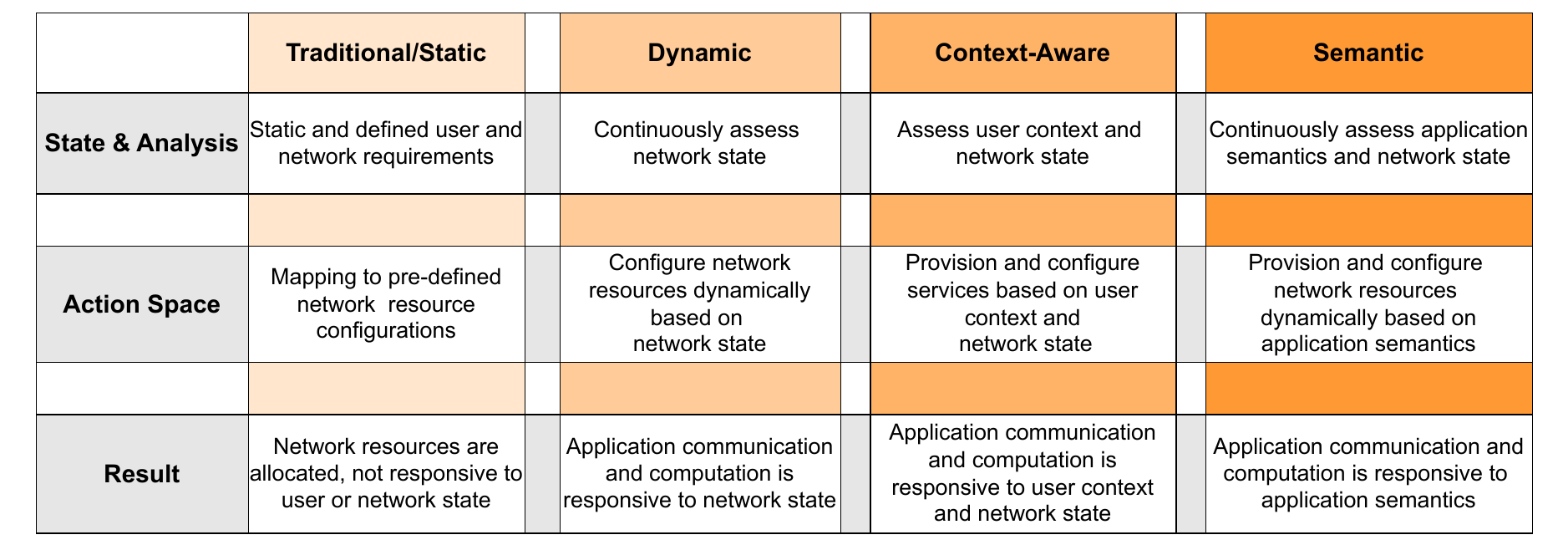}%
   \caption{Comparison between traditional/static resource provisioning, dynamic network slicing, context-aware communication, and proposed semantic slicing.}
  \label{fig:comparison}
\end{figure*}

Applications for \textbf{first responders} need prioritized data transmissions for users such as police, firefighters, and medical personnel. Such prioritization requires various service classes from underlying networks, including mission-critical \ac{PTT}, live video streaming, critical data sharing (image/text/video), \ac{D2D} communication for collaboration within or between teams, and real-time location tracking.
To illustrate the benefits of semantic slicing, let us consider a scenario where a police officer pursuing a suspect needs assistance. The officer initially uses \ac{PTT} to contact nearby officers. Further, they might activate their body cameras to stream live video and share their \ac{GPS} locations. Should the application detect any unusual activity in the video stream, it can automatically alert other units without manual intervention from the officer. Moreover, the network will be able to sense the varying tasks (object detection, alert notification, video stream sharing, etc) as the situation unfolds and vary the provided network resources accordingly. In this scenario, the different users under the same application/vertical, or even the same user under different circumstances might require very different \ac{QoS} and hence might be required to move from one slice to another. A semantic slicing framework will analyze the semantics of the user application data captured by the devices used by first responders and, based on it, take decisions like slice-switching, slice-level resource allocation, etc. 

Semantic slicing can also be useful for \textbf{telehealth} applications, that encompass various service classes such as remote patient monitoring, wearable sensors, tele-consultations (via voice or video calls), remote surgery, diagnostics, data sharing (reports and prescriptions), and emergency telemedicine services.
For instance, chronic patients receiving in-home treatment can be monitored through body-mounted wearable sensors connected to a network. In the event of rapid and critical fluctuations in body parameters, the telehealth system needs to automatically send emergency alerts to the hospital and may initiate a teleconsultation with a doctor for remote diagnostics. Depending on the situation, a smart ambulance equipped with reliable connectivity might be dispatched. This scenario underscores the need for an automated and dynamic semantic slicing framework that can detect any change in the criticality of the patients and provision resources to the slices accordingly so as to minimize disruption in telehealth services due to inadequate communication network resources.

Semantic slicing can facilitate better data security, specifically data access controls via isolation or segregation of data transmissions, crucial for ensuring that sensitive or confidential data streams are accessible only to authorized parties. This is particularly important in sectors such as \textbf{financial services} where data privacy is paramount. 
Also in industrial environments, such as on a production line involving multiple experts, semantic slicing can provide prioritized access to critical data specifically relevant to each expert's tasks. This capability extends to scenarios where infrastructure supports multiple organizations, highlighting the potential of isolation and prioritization to deliver the right data at the right time, thereby enhancing operational efficiency and security. 
This is also relevant in an educational setting with telepresence, where a teacher might have access to additional data that students cannot view. 

Further, in \textbf{telepresence} scenarios, semantic slicing can also enable dividing data into subsets whose transmission and processing are separated making it possible for also the users themselves to have fine-grained control over particular subsets, providing a basis for consent management.


\section{Related Works}\label{sec:related}

To further motivate and clarify the contributions of this work, we expand on recent works and related domains including traditional service provisioning, \ac{DNS}, \ac{AI}/\ac{ML} based network slicing, context-aware communication, as well as semantic communication in general. In Fig. \ref{fig:comparison}, we present a comparative analysis to differentiate between the proposed semantic slicing framework and other relevant existing technologies, specifically tradition/static resource provisioning, dynamic network slicing, and context-aware communication.

Resource provisioning in traditional networks is broadly static 
\cite{static_commag17}, as suggested by \ac{3GPP} in the initial trials of network slicing \cite{3gpp_ts22101}. Usually there are pre-defined user requirements that inform how network resources are provisioned across slices. Hence the network does not react to the dynamically changing network state or user requirements. 
 
\ac{AI}/\ac{ML} methods have recently been extensively used for network slicing and service provisioning. These methods assess the network state and then provision and configure network services accordingly
\cite{AISlice5g_TCCN20}. A \ac{RL} based slice admission and slice congestion control scheme was proposed in \cite{AISlice5g_TCCN20}. An \ac{AI} native slicing framework was proposed in \cite{AINative_NS6G_WCMag22} where different related use cases and open research challenges were also talked about. In this paper, the authors discussed how \ac{AI} can take a central role in preparation (demand prediction, slice admission), planning (placement of \acp{VNF}, reservation of resources), operation (orchestration of resources, selection of \ac{RAT}) of \ac{NS} and thereby backed the need for an \ac{AI} native slicing framework. A comprehensive survey of existing works related to \ac{AI}/\ac{ML}-assisted network slicing is provided in \cite{survey23_ml_ns}. In these works, various \ac{ML} algorithms like neural networks, support vector machines, deep Q learning etc. were leveraged in solving challenges pertaining to three different phases of the \ac{NS} life-cycle: preparation, commissioning, and operation. In \cite{survey23_ml_ns}, it was also highlighted that most of the existing works are simulation-based and moving from theory to practice or real-world experimentation in this domain is still an open research problem.  




\ac{DNS} was addressed in several works that focused on dynamically scaling up or down resources allocated for each slice according to the varying demand \cite{DNS_survey_Springer20}. This work talks about the main enablers of \ac{DNS} like \ac{SDN}, \ac{NFV}, cloud infrastructure, \ac{E2E} orchestration etc. It also emphasizes the need for continuous monitoring and analyzing the network in order to manage resources effectively.
Further, a queuing theory-based dynamic auto-scaling method was proposed for network slice orchestration in \cite{DNS_E2E_Taleb_TMC20}. Dynamic slicing for individual network components like \ac{RAN}, edge, \ac{CN}, or \ac{TN} have also been considered. For example, a \ac{DNS} scheme for multi-tenant heterogeneous cloud \ac{RAN} is proposed in \cite{DNS_CRAN_TWC18} to maximize the number of users served. 
In \cite{DNS_Edge_DRL_GC20}, a Q-Learning based dynamic edge slicing technique is proposed to optimize the policy to admit slice requests. Whereas, in \cite{DNS_scale_Access20}, a multivariate time-series based forecasting of \ac{VNF} resources was proposed for \ac{CN} slicing.

Context-aware communication \cite{context_comm_WC02},
on the other hand, senses the environment precisely to assess user context (e.g., at home or at work, busy or available, mobile or stationary), and configure applications based on that context. As a result, application communication and computation become sensitive to user context, allowing for finer-grained control of user \ac{QoE}. 
Very few works have considered context-awareness in network slicing for the purpose of resource allocation, management, or orchestration. Wu et al. \cite{context_icnc20} used a genetic algorithm based optimization for context-aware resource allocation based on user-specific customized requirements like screen size, age etc. \cite{context_IoTJ22} considered context-awareness in user authentication based on user context information like location, timestamp etc.
However, in these works, the resource provisioning is based only on user context, disregarding the semantic content of the sensed data, which is expected to give a bigger picture of the overall application and environment, and build a situational awareness that goes beyond the need of individual users. 


Resource allocation and \ac{NS} for specific applications have also been examined. In \cite{DNS_IoT_conf18}, authors proposed to dynamically scale in or scale out the number of resource blocks based on the traffic requirement of \ac{IoT} applications. A multi-agent \ac{RL} based dynamic slice resource allocation was proposed for vehicular networks in \cite{DNS_Vehi_TITS24}. Most recently, intent-based network slicing has received more attention (see, e.g., 
\cite{Intent_eucnc21}), 
in which slicing is based on high-level requirements, called ``intents", while how to achieve those requirements (say, policy) is left open. 
But none of these frameworks consider application data semantics for decision making related to slice resource allocation.



Semantic communication is one of the futuristic technologies that can help go beyond the limitations of conventional communication systems and can potentially transform existing networks \cite{lessdata_walid_arxiv22}. A \ac{FL} based edge intelligence was proposed in \cite{CommMag21_SemAwareNetwrkng} for supporting a resource-efficient architecture towards semantic-aware networking. \ac{AI}/\ac{ML} based methods like neuro-symbolic \ac{AI}, causal representation learning, generative \ac{AI} has been explored for semantic modeling \cite{neurosymb_walid}. 
\cite{Task_MU_SemCom_JSAC22} considered task-oriented multi-user semantic communication where three different frameworks are proposed for specific tasks like image retrieval, visual question answering, and machine translation. Very few works like \cite{PerfOptSC_JSAC22} have considered cellular network infrastructure while evaluating the performance of semantic communication, let alone a network slicing framework that considers semantics. Recently, a semantic edge slicing application was proposed, which allocates resources to deep learning services involving resource-expensive tasks~\cite{sem-o-ran-infocom23}.

This article builds upon a recent study~\cite{loven2023semantic} by, in part, the same team, which presented a general overview of the idea of semantic slicing. Discussing the topic in more detail, this article describes potential use cases, takes a more in-depth look at related works, and delves deeper into the implementation aspects, proposing a novel architecture to enable semantic slicing on 6G networks. 

\section{Network Slicing in 5G Service Based Architecture}
\label{sec:5gslicing}

\begin{figure}
   \centering
   \includegraphics[width=\columnwidth]{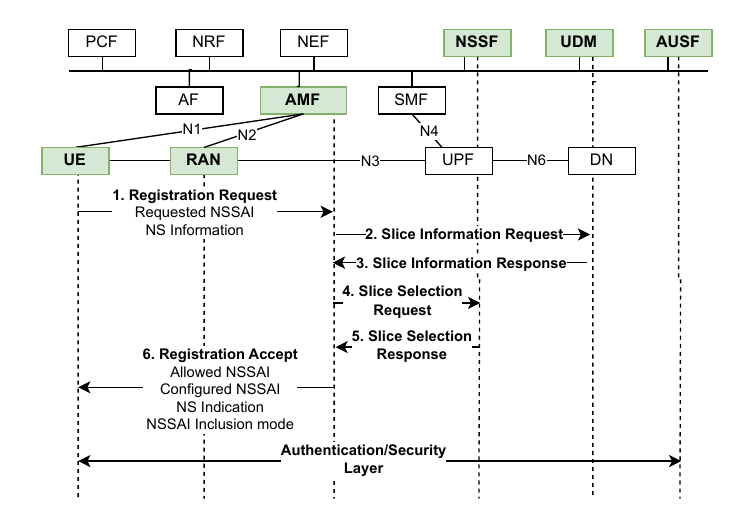}%
   \caption{Signaling involved in 5G Network Slicing}
  \label{fig:ns_timing}
\end{figure}

In this section, we briefly describe the working principle of network slicing in 5G, and then in the following section, we articulate how the existing architecture can be upgraded to the proposed semantic slicing framework. The core entities involved in \ac{NS} are (i) \textit{Tenant} i.e. consumer of the \ac{NS} services, (ii) \ac{NSP} i.e. the entity which provides \ac{NS} as a service (e.g. \acp{MVNO}), and (iii) \ac{NSA} an entity present at the premises of infrastructure provider who has a complete picture of the available resources and has control over how to use and manage them. A tenant goes to the \ac{NS} with a set of communication service requirements that are specific to its business applications. This can be done through a \ac{NST} which lists a group of attributes that define the corresponding slice. Once the \ac{NSP} receives the service requirements its job is to assess the \ac{E2E} NS requirements and delegate to the subnets like \ac{RAN}, \ac{CN}, and \ac{TN} accordingly.   

Fig. \ref{fig:ns_timing} shows the interaction between different \acp{VNF} (highlighted in green) involved in network slicing in the \ac{SBA} of 5G. It also includes a timing diagram showing the different steps involved in the attachment process of a user, called \ac{UE} in 5G terminology, to a slice.
The \ac{NSSF} was introduced in the 5G \ac{CN} in order to take the central role in the selection of slices for \acp{UE}.
The network uses \ac{NSSAI} to identify which slice the \ac{UE} belongs to. \ac{NSSAI} is a combination of up to eight \acp{S-NSSAI} each of which refers to the different slices the \ac{UE} can be served by at one time. The \ac{S-NSSAI} is a combination of two fields namely (i) \ac{SST} and (ii) \ac{SD}. Different service profiles like \ac{uRLLC}, \ac{mMTC} or \ac{eMBB} can correspond to different combinations of \ac{SST} and \ac{SD} values.

During \ac{RRC} connection request and \ac{NAS} registration procedure, the \ac{UE} sends the \ac{S-NSSAI} i.e., the tuple of \ac{SST} and \ac{SD}. Once the request is received by the \ac{AMF} it shares the details with \ac{UDM} and receives slice information as the response. Subsequently, \ac{AMF} sends the slice selection request with \ac{NSSF} to enquire if the \ac{UE} is allowed to connect to this slice. Once \ac{AUSF} is finished with slice authentication/security-related procedures, \ac{AMF} sends a registration accept message to the \ac{UE}.

Management of \ac{NS} in the infrastructure of a network operator is performed primarily through three functions defined by the \ac{3GPP}: \ac{CSMF}, \ac{NSMF}, and \ac{NSSMF}.
The \ac{CSMF} is responsible for receiving the communication service-related requirement from the \ac{CSC}, translating them to \ac{NS} related requirements, and passing them to the \ac{NSMF}. Once the \ac{NSMF} receives the \ac{NS} related requirements, its job is to create the \acp{NSI} based on them. It also extracts the \ac{NS} subnet related requirements and shares them with the \ac{NSSMF}.

The high level functions performed by the different entities in \ac{NS} are the following \cite{etsi2018ngp011_ns} :

\begin{enumerate}[(i)]

\item  \textit{Network slice subnet discovery function} - establishing central repository of resources exported by \acp{NSA} to \ac{NSP}

\item \textit{Network slice subnet augment function} - addition/deletion of resources to the slices based on the requirement, in the event of a change in one or more parameters of the provided service, e.g., delay budget changing from 10 ms to 5 ms.

\item \textit{Network slice mapping function} - takes node and edge segment as input from aggregated resource database which is output of computation function and  associates infrastructure elements to logical elements in \ac{NSI}

\item \textit{Resource computation function} - to identify which resources and paths are suitable and a certain service graph representing a network slice or service

\item \textit{Network slice delegation function}- after resource computation and mapping are successfully done, this block delegates them to subnets 

\item \textit{Report aggregation function} - collects and consolidates reports corresponding to performance monitoring of the \ac{E2E} framework

\item \textit{Service assurance function} - takes appropriate steps to ensure all the slices are getting services assured as part of the \ac{SLA} including overall monitoring by the \ac{NSA} and each subnet-level monitoring by respective \acp{NSA}

\item \textit{Tenant operated network service function} - to give more control to tenant over the resources specific to its slice and follows this flow of interface: tenant $\rightarrow$  \ac{NSP}  $\rightarrow$ \ac{NSA} $\rightarrow$ network.
\end{enumerate}

In Fig. \ref{fig:semslice-arch} we highlight in blue these functions and how they act as a medium of interaction between different blocks of \ac{NSP} and \ac{NSA}. With this background about the current \ac{3GPP} standard for \ac{NS} architecture, in the following section, we describe in detail the proposed semantic slicing framework. 

\section{Proposed Semantic Network Slicing Framework}\label{sec:semantic}

\begin{figure*}
\centering
\includegraphics[width=\textwidth]{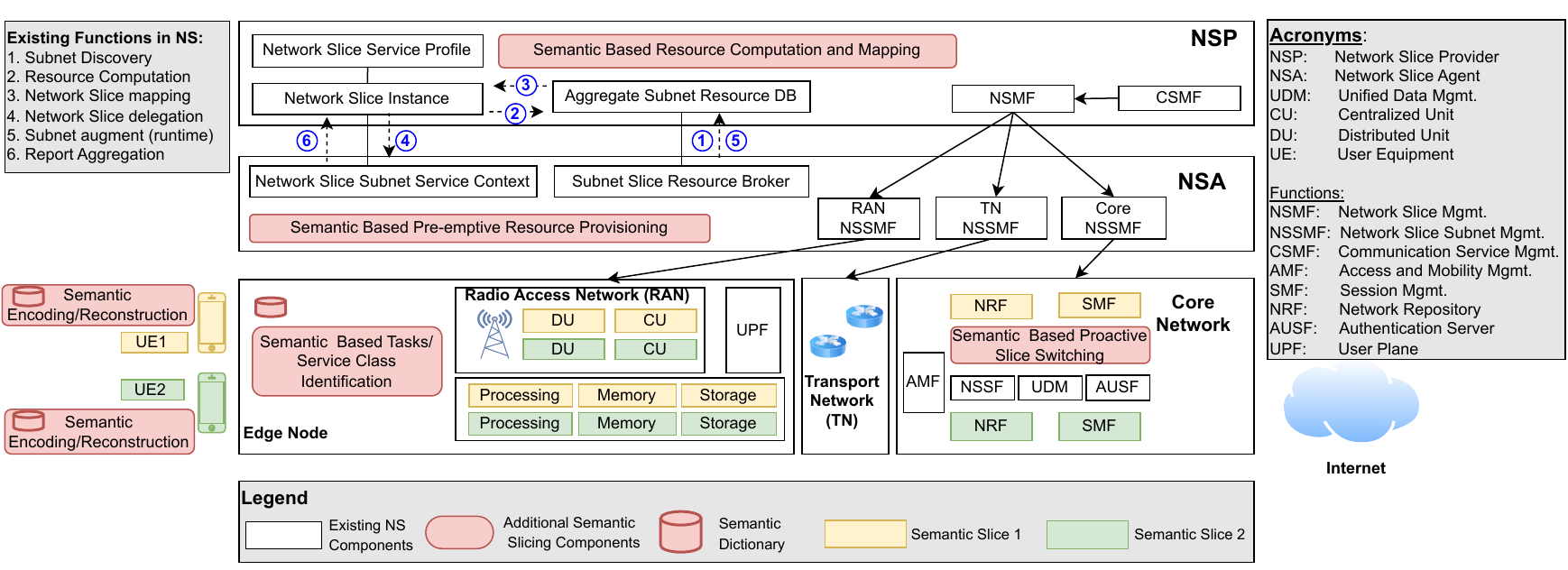}
\caption{Proposed Semantic Slicing Architecture. The additional components related to Semantics are highlighted in red. The \acp{VNF} and resources dedicated to two different slices are highlighted in yellow and green. \ac{UE}1 and \ac{UE}2 belong to Slice1 and Slice2 respectively.}
\label{fig:semslice-arch}
\end{figure*}


Fig. \ref{fig:semslice-arch} shows the architecture of the proposed semantic slicing framework. Next, we explain in detail each of the additional building blocks or functional components that are envisioned to be added to the existing \ac{5G} \ac{NS} architecture in order to enable the proposed semantic slicing framework.


\subsection{Semantic Encoding/Decoding}

We assume that all the users involved in the proposed framework support semantic communication by default. Therefore it is fair to assume that each user has semantic encoding and reconstruction blocks. If they do not have sufficient computation capability on board, nearby edge devices can be provisioned in the slice by the orchestrator. While generating the semantic representation of the data, it has to be kept in mind that the same will be used later to take decisions related to network slicing. Therefore, it is not sufficient to create a minimalistic semantic representation instead, it has to be a meaningful one that may aid in reasoning and intelligent orchestration of slice resources. Each \ac{UE} can belong to different applications or verticals and hence might be exposed to very different types of data. All \acp{UE}s have access to a common dictionary or knowledge base. This dictionary keeps on being updated as part of a continuous learning process. 

Let $X$ denote the raw data stream to be semantically encoded, and let $Y$ be the semantic encoding of $X$ under dictionary $D$, and the semantic mapping function $f^S_D$. Also, let $\mathbf{X}$ be the set of all possible input data. For the proposed semantic slicing approach, meaningful information that can support the orchestration of resources can be linked to what types of computational tasks are needed. Hence, let $\mathbb{T}$
represent the set of all computational tasks  $T_i$ 
identified in $X$, 
via its semantic representation $Y$, using some semantic computational and reasoning method $Z^S_D$. Variables $X, Y, T_i$ are all time-dependent, however, we drop the time index here for clarity. Then we can abstract the semantic encoding and task extraction as follows

\begin{equation}
    Y = f^S_D(X), \text{where } X \in \mathbf{X} , 
\end{equation}

\begin{equation}
    T_i(X) = Z^S_D(Y), \text{where } T_i \in \mathbb{T} \quad \forall i \in  \{1, \ldots, n\}
\label{eq_task}
\end{equation}

We expand on the mapping and semantic reasoning in the following subsections.

\begin{figure*}[!htb]
   \centering
\includegraphics[width=0.8\textwidth]{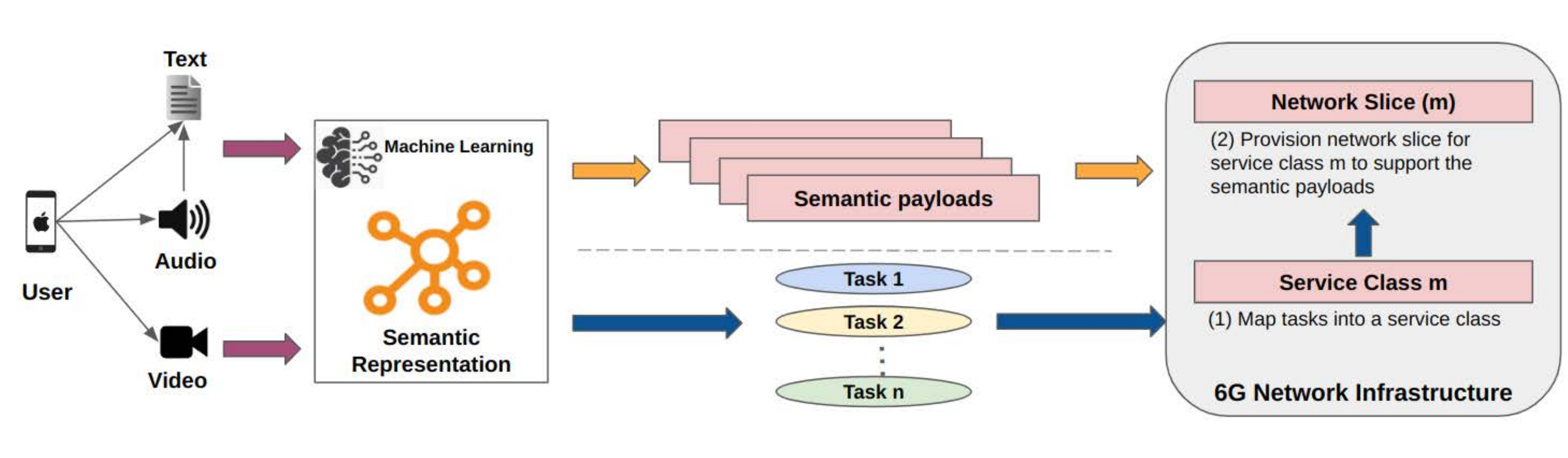}

   \caption{Service Class Identification}
  \label{fig:servclass}
\end{figure*}

\subsection{\Ac{SSCI}}
\label{sec:sem_service}
In semantic communication, the \acp{UE} utilizes some algorithm for semantic representation of the data it needs to transmit (some examples are given in Section \ref{sec:sem_repr}). In order to provide an appropriate network slice to the \ac{UE}, the network has to detect which service class it belongs to. Hence, it needs to detect semantic events and meaning, from the variation in semantic representation of the data, which will help the network understand the change in network service requirements. This is accomplished through a semantic-based task identification functional block, which is shown in Fig. \ref{fig:servclass}, and mathematically defined in (\ref{eq_task}). Once tasks are identified, then mapping to a service class is processed, and the selected service class is passed on to the rest of the functional blocks for making decisions related to slice switching, slice resource computation and mapping, resource provisioning etc., which are described in the following subsections.

\subsection{Catalog of generic service classes}

The tenant of the network slice can have users of diverse service classes. For example, a tenant that provides telehealth services may need \ac{eMBB} slice for remote patient monitoring. It might also need \ac{uRLLC} slice for alert notification to doctors in case any rapid fluctuation in the body parameters of the patient is detected. Therefore a catalog of generic service classes along with their respective \acp{KPI} is to be maintained. These \acp{KPI} are decided based on mutual agreement between the tenant and the \ac{NSP} and are part of the \ac{SLA}. An example of such a catalog of generic service classes for a specific use case application is provided in Table \ref{tab:task-qos-fr} in Section \ref{sec:usecase}. As can be seen from the table, for a specific task, for example Event Detection, that is identified in the use case application there is a set of \ac{QoS} metrics that are needed to support this task, such as compute resources, delay sensitivity, etc. 

Depending on the task $T$, the network needs to identify the service class that best serves task $T$ based on the required 
\ac{QoS} metrics. Let $C_i$ denote the service class corresponding to Task $T_i$. Given a specific task $T_i$, then the corresponding service class can be described in terms of defined \ac{QoS} metrics. let $W$ be a function that maps the task to a vector representing the quantitative \ac{QoS} requirements in terms of a defined set of metrics. Assuming that the values of ${QoS}_i$ are non-negative integer numbers (i.e. $\in \mathbb{N}$), then the mapping can be abstracted as follows:
\begin{align}
    C_i(T_i)  = W(T_i) = & [{QoS}_1, \ldots, {QoS}_k ] \nonumber \\
    {\text{ where }} &  \forall  T_i \in \mathbb{T}{\text{, }}  \exists{\text{  }}   {QoS}_i \in \mathbb{N}   
\label{eq3}
\end{align}


\subsection{\Ac{SPSS}}
The semantic slicing orchestrator in the proposed framework will provide an initial default slice to each \ac{UE} which is enough for the \ac{UE} to extract the semantic features of the information to be sent. Once the semantic features are extracted and the corresponding service class is identified, by \ac{SSCI} as described in Section \ref{sec:sem_service}, users will be switched to the appropriate slice based on their respective \ac{QoS} requirements. The \ac{SPSS} will monitor the service class identified by \ac{SSCI} and whenever there is a change in the same, it will initiate a slice switching request. The slice orchestrator will decide whether to change the current slice of the \ac{UE} to the requested slice or not based on different factors like current load on the slice, additional resource required by the slice to serve the current \ac{UE} etc.



\subsection{\Ac{SRCM}}
One of the most complex functionalities in the \ac{NS} framework is the resource computation for each of the slices. There are two types of resources: (i) Link and (ii) Node. While the former has traffic and path-related constraints, the latter concerns compute or storage-related constraints. Semantic models can represent user context and help understand the service requirements. Based on the semantic model of the application data of each user in a slice, the \ac{NS} orchestrator computes the resource required to satisfy the respective \ac{QoS} requirements. Primarily, a functional mapping is obtained between resources and the \acp{KPI} of each of the slices. Subsequently, the physical resources of the network are mapped to the logical ones. While performing the resource computation and task mapping in the proposed semantic slicing framework, the communication-computation trade-off has to be considered. The more concise the semantic representation is, the less will be the communication resources required to transmit them to the receiver side. However, more compute resources might be required in semantic encoding i.e., to convert the actual data to its semantic representation, or in semantic reconstruction i.e., to get the actual data back on the receiver side.

\subsection{\Ac{SPRP}}
Once the resource computation and mapping are complete, it is the job of the network slice orchestrator to provision resources based on the same. The semantic models of application data of all users also give the orchestrator an idea about what is going on in a particular service area. For example, in a first responder application, if the semantic model indicates some incident has happened, it means more and more users of that area might require mission-critical communication service. Hence, the network orchestrator can preemptively add more resources to the critical slice to avoid any possible disruption in service. Similarly, once everything settles down and comes back to normalcy, that should be reflected in the user data. Therefore, the orchestrator can remove resources from critical slices and give back to best-effort slices.  


\section{Use case Scenario - First Responder}\label{sec:usecase}

\begin{figure*}[!htb]
   \centering
   \includegraphics[width=0.8\textwidth]{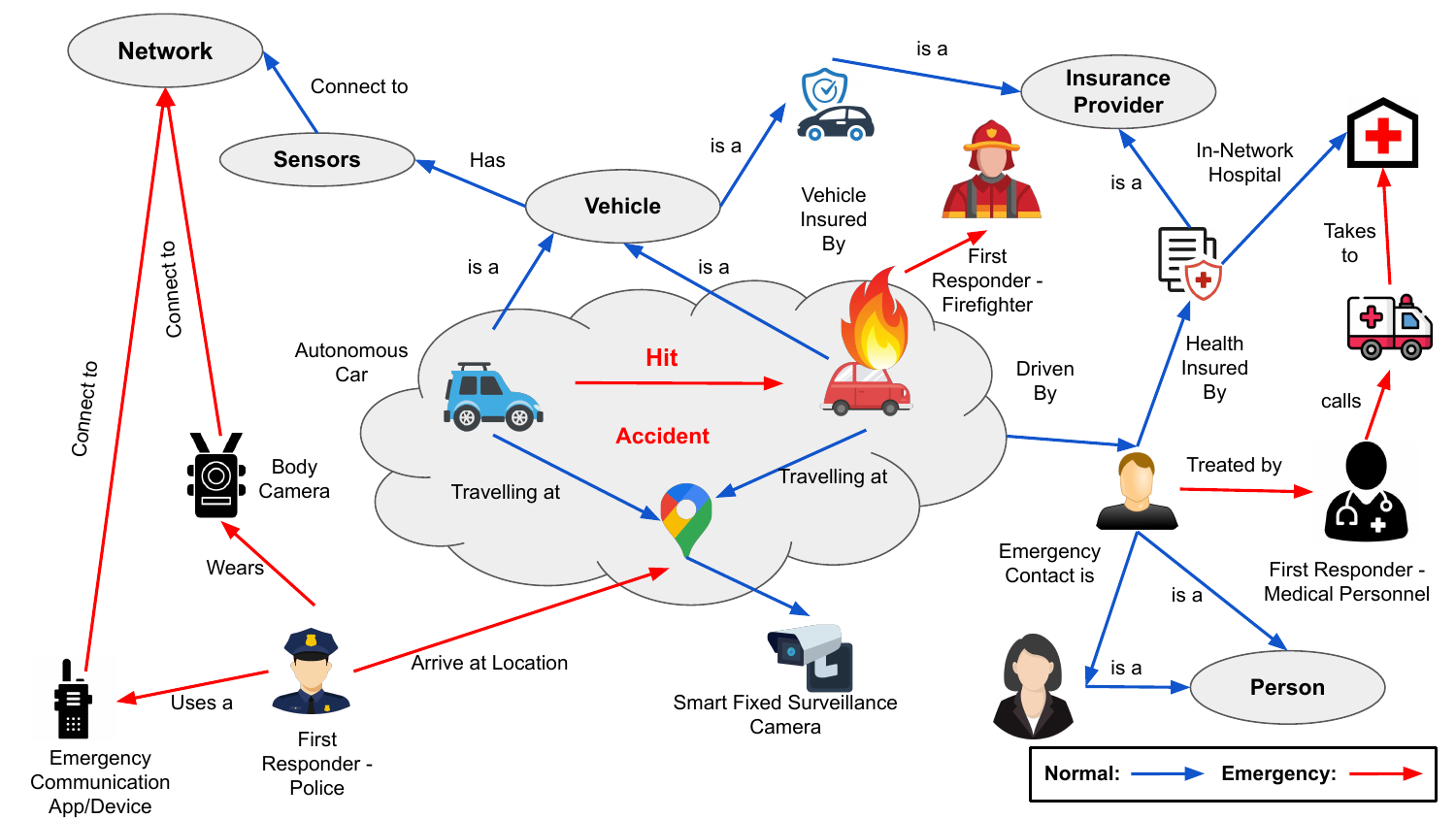}%
   \caption{Knowledge Graph based semantic representation of a First Responder scenario}
  \label{fig:kg_combined}
\end{figure*}

Let us consider the example of a disaster scene that needs attention from first responders, who are the first to arrive at the scene whenever there is an accident, medical emergency, or any other public safety issue, such as an earthquake. \Ac{SEMA} can be considered to be a tenant that is under contract with an \ac{NSP} for the network slicing services. Different first responders like police, firefighters, or medical personnel can be considered as users under \ac{SEMA} i.e. the tenant. The \ac{SEMA} will be subscribed to one or more slices and users will utilize the resources allocated to each of those slices as per their individual requirements.


The first responders, i.e. \ac{NS} users, can be involved in different tasks like continuous monitoring, object detection, event detection, alert notification, tracking an object of interest, telehealth, remote control for emergency operations, etc. Each of these operations can act as a trigger to the other, whereas each of these may require a very different quality of service. For example, continuous monitoring of a public area through video surveillance may have a high bandwidth \ac{QoS} requirement, while object or event detection may require high compute power at the edge, and so on. 



To further illustrate the above concepts, consider the scenario in which some persons are under surveillance and object tracking. When a serious injury or death event is detected by the monitoring service, this should trigger an alert service to send a high-priority notification to concerned authorities or first responders. The alert notification likely requires \ac{uRLLC} \ac{QoS}. Now, after the police or law enforcement officers reach the scene of the emergency, they might need to notify medical personnel if their attention is required. Based on the condition of the victims, telehealth services or remote support might be required. Also firefighters might be required to extinguish fire caused by collision or engage in some other rescue/relief operation. These are expected to require high bandwidth, low latency, and ultra-reliable network links. Furthermore, when a large number of first responders reach the spot, for them to communicate with each other or to communicate to distant control rooms, massive user support might be required.


Therefore, if an organization (e.g. \ac{SEMA}) as the tenant of \ac{NS} has subscribed to first responder services, it might require a very diverse set of \ac{QoS} at different points in time. Hence, the organization users might be required to be associated with different slices over time. Table \ref{tab:task-qos-fr} summarizes different tasks involved in applying first responder services and their respective \ac{QoS} requirements. The network users are classified into one of the service class categories in Table I based on their tasks and then may be assigned slices accordingly. For simplicity, we consider discrete levels of \ac{QoS} like HIGH, \ac{AVG}, or LOW in Table \ref{tab:task-qos-fr}. Based on the requirements of the applications to be served, more fine-grained levels of \ac{QoS} can be used by the \ac{NSP}. In \cite{semantic_kpis_access23}, authors surveyed some metrics that can be used to evaluate the performance of semantic communication systems. These can also act as a baseline for additional \ac{QoS} parameters while defining the \ac{SLA}s of the proposed semantic slicing framework.


\begin{table*}
\caption{\ac{QoS} Requirements of different tasks related to First Responder application}
\label{tab:task-qos-fr}
\centering
\begin{tabular}{|l|lllllll|}
\hline
\multicolumn{1}{|c|}{\textbf{Tasks}}     & \multicolumn{7}{c|}{\textbf{QoS Parameters}}  \\ \hline
\textit{}                                & \multicolumn{1}{l|}{\textit{\textbf{Bandwidth}}} & \multicolumn{1}{l|}{\textit{\textbf{Delay Sensitivity}}} & \multicolumn{1}{l|}{\textit{\textbf{Reliability}}} & \multicolumn{1}{l|}{\textit{\textbf{Scale}}} & \multicolumn{1}{l|}{\textit{\textbf{Compute}}} &  \multicolumn{1}{l|}{\textit{\textbf{Storage}}} & \textit{\textbf{Mobility}} \\ \hline
Continuous Monitoring                    & \multicolumn{1}{l|}{HIGH}                        & \multicolumn{1}{l|}{LOW}                                 & \multicolumn{1}{l|}{AVG}                           & \multicolumn{1}{l|}{AVG}                     & \multicolumn{1}{l|}{AVG}                            & \multicolumn{1}{l|}{HIGH} & LOW                        \\ \hline
Object Detection                         & \multicolumn{1}{l|}{AVG}                         & \multicolumn{1}{l|}{AVG}                                 & \multicolumn{1}{l|}{AVG}                           & \multicolumn{1}{l|}{AVG}                     & \multicolumn{1}{l|}{HIGH}   & \multicolumn{1}{l|}{HIGH}         & LOW                        \\ \hline
Event Detection                          & \multicolumn{1}{l|}{AVG}                         & \multicolumn{1}{l|}{HIGH}                                & \multicolumn{1}{l|}{AVG}                           & \multicolumn{1}{l|}{AVG}                     & \multicolumn{1}{l|}{HIGH}
& \multicolumn{1}{l|}{AVG} & LOW                        \\ \hline
Alert Notification                       & \multicolumn{1}{l|}{LOW}                         & \multicolumn{1}{l|}{HIGH}                                & \multicolumn{1}{l|}{HIGH}                          & \multicolumn{1}{l|}{HIGH}                    & \multicolumn{1}{l|}{LOW} & \multicolumn{1}{l|}{LOW}                            & LOW                        \\ \hline
Tracking object of interest              & \multicolumn{1}{l|}{LOW}                         & \multicolumn{1}{l|}{AVG}                                 & \multicolumn{1}{l|}{AVG}                           & \multicolumn{1}{l|}{AVG}                     & \multicolumn{1}{l|}{HIGH} & \multicolumn{1}{l|}{HIGH}                           & LOW                        \\ \hline
Tele-health                              & \multicolumn{1}{l|}{HIGH}                        & \multicolumn{1}{l|}{HIGH}                                & \multicolumn{1}{l|}{HIGH}                          & \multicolumn{1}{l|}{AVG}                     & \multicolumn{1}{l|}{AVG} & \multicolumn{1}{l|}{HIGH}                            & LOW                        \\ \hline
Remote control                           & \multicolumn{1}{l|}{HIGH}                        & \multicolumn{1}{l|}{HIGH}                                & \multicolumn{1}{l|}{HIGH}                          & \multicolumn{1}{l|}{LOW}                     & \multicolumn{1}{l|}{AVG} & \multicolumn{1}{l|}{LOW}                            & LOW                        \\ \hline
Push to Talk (PTT) & \multicolumn{1}{l|}{LOW}                         & \multicolumn{1}{l|}{AVG}                                 & \multicolumn{1}{l|}{AVG}                           & \multicolumn{1}{l|}{HIGH}                    & \multicolumn{1}{l|}{LOW} & \multicolumn{1}{l|}{LOW}                            & HIGH                       \\ \hline
Smart Ambulance                          & \multicolumn{1}{l|}{HIGH}                        & \multicolumn{1}{l|}{AVG}                                 & \multicolumn{1}{l|}{AVG}                           & \multicolumn{1}{l|}{AVG}                     & \multicolumn{1}{l|}{HIGH} & \multicolumn{1}{l|}{HIGH}                           & HIGH                       \\ \hline
\end{tabular}
\end{table*}

In Fig. \ref{fig:kg_combined}, a prospective semantic representation of a typical accident site, which needs attention from first responders, is shown. In this case, we utilize Knowledge Graphs for the semantic representation, but other algorithms such as causal representation learning or \ac{ML} based semantic feature extraction can also be used for this purpose. 
In Fig. \ref{fig:kg_combined}, when everything is normal, road traffic is monitored by a \ac{CCTV} camera, which is connected to the 5G network and is monitored by law-enforcement officials under \ac{SEMA}.
In this example use case, let us assume an accident occurs where a bus hits a car, which gets damaged, and the car driver gets badly injured. Due to the collision, the bus then catches fire and needs the firefighters to come and extinguish the fire. The injured car driver must be tended to on-site possibly and then taken to a nearby hospital, and the car needs to be repaired. Both the health insurance provider and the car insurance provider must be contacted. All these activities will require intervention from different first responders and will also have varied communication service requirements. The proposed semantic slicing framework will be able to analyze the semantic representation in the computer continuum under normal and emergency situations and dynamically identify and provision the network slices with appropriate resources to each user. The resources associated with the different slices will be provisioned based on the knowledge inferred from the data semantics. For example, when the system detects that the accident has occurred and detects possibly injured personnel, then it requires several medical personnel to be on-site and possibly some telehealth scenarios. Hence, the communication network slices will be assigned and resources will be coordinated accordingly. When a massive fire has broken out, and several firefighters are needed at the location to control the fire, the communication network will increase the amount of resources to the corresponding slice. 





\section{Initial Results}\label{sec:init_results}

\begin{figure}[!htb]
   \centering
   \includegraphics[width=\columnwidth]{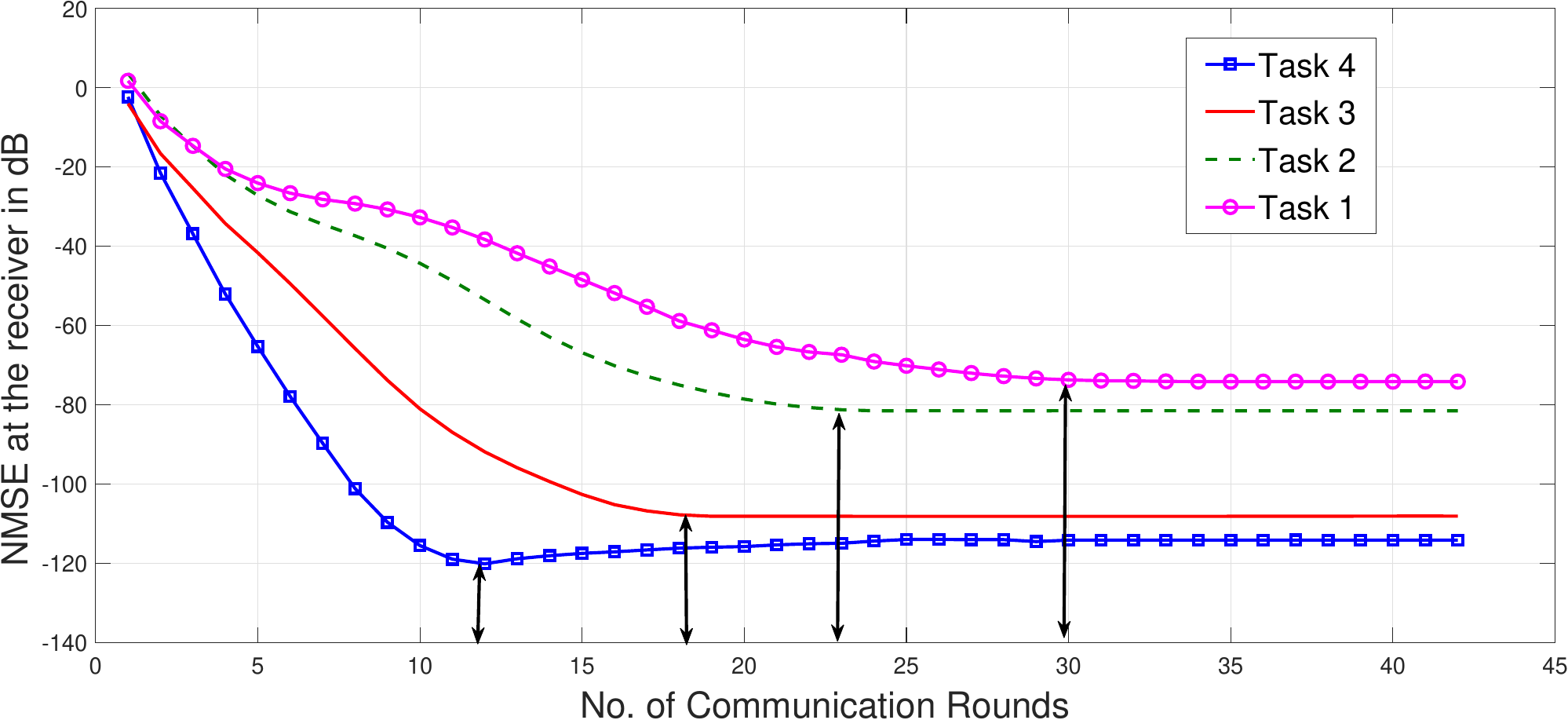}%
   \caption{Convergence of single-link semantic communication system~\cite{CHRISTO00}}
  \label{fig:res}
\end{figure}

In this section, we discuss a simple single transmitter and single receiver use case that will highlight how encoding and decoding in a semantic communication scenario can improve the overall network performance. This can be a single-link basis for our broader semantic slicing vision. This example is based on our results in \cite{CHRISTO00}. Specifically, in \cite{CHRISTO00}, we considered a scenario in which a semantic transmitter extracts the semantic content elements that precisely represent its input. The extracted semantic content elements follow a \ac{SCM} that is essentially a directed acyclic graph that captures the \emph{cause and effect relationship} among a set of variables. Here, the \ac{SCM} represents the relationships among the semantic content elements of the data. Those elements are learned through the framework of generative flow networks (GFLowNets)~\cite{GFLOWNET02}. To enable the semantic transmitter and receiver to communicate, we build a semantic language based on the extracted \ac{SCM}. The design of the language itself is posed as a signaling game, and we show that the equilibrium of this game enables effective encoding of the semantic content elements into a minimalistic language whose space is much smaller than that of the original set of content elements. Using symbolic AI, we show that this language can be compositional and, thus, it enables effective single-link semantic communication. We then analytically prove~\cite[Theorem 2, Theorem 3]{CHRISTO00} that the designed semantic communication system reduces the amount of information circulating over the network and achieves a higher semantic reliability compared to classical transmission. In Fig.~\ref{fig:res} from \cite{CHRISTO00}, we show the convergence behavior of the proposed semantic system as time evolves for different sequentially arriving tasks that represent variations in the input space. Here, we use the \ac{NMSE} as a measure that represents the effectiveness of the action selection of the receiver based on the transmitted causal information. In this figure, the term ``task'' is used to represent a set of logical formulas, each being a mapping from the space of the language vocabulary to the logical conclusions that the system can make. A larger task index indicates a ``harder'' task for the system. From this figure, we observe that task 1 needs 30 rounds to converge, while task 4 needs only 10 rounds. This demonstrates that the use of a semantic language can, over time, lead to a faster execution of tasks, because of the ability to acquire knowledge through the combination of the \ac{SCM} model, GFlowNets, and symbolic \ac{AI}. This acquired knowledge means that, over time, the semantic language can rapidly adapt to new tasks, and its learning overhead significantly decreases. This approach, when generalized across the network, enables each communication link to intrinsically build a language-based knowledge base (dictionary), that can then be used to make transmission decisions and, more broadly, networking slicing decisions. 

\section{Open Challenges} \label{sec:challenges}
To fully realize the proposed framework, more research and development is needed. In this section, we identify and expand on select open research problems in semantic slicing.

\subsection{Semantic Representation} \label{sec:sem_repr}
In exploring the representation of semantic information, the objectives are twofold: to deduce meaning from signals and to encapsulate this meaning in models that are both rich in information and low in computational complexity. However, achieving these goals is not without challenges. In the following we discuss some prospective techniques which can be used to extract semantic representation from the application data.

\subsubsection{Causal representation learning}
Causal representation learning enables deriving meaningful state representations from event data logs through process mining. While powerful in deducing causal relationships between events and states, it faces challenges in handling the vast volume of data generated by sensors and systems. The complexity of accurately identifying patterns and constructing models that represent underlying processes can be computationally intensive, requiring sophisticated algorithms and substantial processing power.

\subsubsection{Ontology and Knowledge Graphs} 
These tools provide a structured approach to semantic information modeling. Ontologies and knowledge graphs, despite their effectiveness in representing complex interrelations and supporting semantic querying, struggle with the dynamic nature of real-world environments. Keeping these models updated and relevant in rapidly changing scenarios requires continuous revision and can lead to scalability issues. Furthermore, integrating disparate knowledge sources into a cohesive graph presents significant challenges in ensuring consistency and accuracy.

\subsubsection{Ontology databases/dictionaries} The semantic encoding and reconstruction will require synchronized and efficient access to ontology databases/dictionaries that are not necessarily previously seen, which could be an opportunity for \ac{AI}/\ac{ML} based approaches.


\subsubsection{Pervasive service composition} The service decomposition will dynamically compose services based on semantic information, aiming for seamless interaction across applications and devices. The challenge in this domain is the heterogeneity of services and the complexity of their integration. Achieving semantic interoperability between disparate services, each with its own data models and protocols, requires sophisticated matching and translation mechanisms.

\subsubsection{Cross-application Semantics} There will be a need to establish a framework for sharing semantic information across applications to enhance interoperability and enable richer contextual understanding. The primary challenge here is the development of standardized models and protocols that can be universally adopted, ensuring consistent interpretation of semantic information. Additionally, maintaining privacy and security while sharing semantic information across applications presents significant hurdles, requiring robust encryption and access control mechanisms.

\subsubsection{Other approaches}
\ac{AI}/\ac{ML} models focusing on perception and \ac{HAR} sensors aim to interpret and represent semantic information from various data sources. The primary challenge here lies in the diversity and variability of sensor data, which can lead to difficulties in model training and generalization. Additionally, distinguishing between relevant and irrelevant data, especially in noisy environments, requires advanced filtering techniques and can significantly impact the computational efficiency of these models. Moreover, \ac{LLM} can map between a language-based description and a semantic representation. However, deriving that language-based description from sensor data, and the efficient implementation of \ac{LLM} mapping thereafter remains a challenge. A promising approach leverages \ac{LLM} to generate the semantic representation in the desired format.

\subsection{\ac{AI} for Slicing and Slicing for \ac{AI}}

Towards the overarching idea of building an AI-Native semantic slicing framework, both \ac{AI} and slicing are expected to complement each other. While \ac{AI} will empower slice management and orchestration with intelligent decision-making, the slicing framework also has to make sure that enough resources are provided to the AI algorithms to run efficiently.

One of the main tasks here is to enable real-time data model creation and support micro-services and data pipelines for intelligent slicing. For streamlining and automating the process of development, deployment, and management of \ac{AI}/\ac{ML} models there is a need to use \ac{MLOps} platforms which provide solutions for the automation of the entire \ac{ML} pipeline, from pre-processing of data to model training and then model deployment. They also enable the following: (i) automatic allocation of GPU and other compute resources for \ac{AI} related tasks; (ii) real-time monitoring of the trained models' performance; (iii) creating an ecosystem for researchers to collaborate by sharing and reusing code, model, and data; and (iv) faster scale-up capabilities to satisfy growing demands.

Adoption of \ac{AI}-driven \ac{CLA} in the proposed semantic slicing framework requires considerations of (i) data coverage
and quality for accurate decision-making; (ii) horizontal (across domains) and vertical (within a domain) 
coordination; (iii) utilization of \ac{AI} models that are more explainable and can be monitored; (iv) utilization of
\ac{AI} models that can perform cross-layer prediction and interaction across network layers; and (v) advanced
monitoring of \ac{CLA} outcomes and performance.

Towards slicing for AI, the pivotal objective is to create multiple flavors of the network from which the most optimal one can be used for implementing an AI algorithm. For example, the networking for a distributed learning framework will be very different compared to a centralized approach. A flavor or instance here means a combination of AI algorithm, architecture of learning, resource requirement etc. In this direction, the latest \ac{ORAN} technology might be looked into, which brings in near real-time and non real-time \acp{RIC} that can provide a platform for network data analytics aiding in intelligent resource orchestration. In this context, allocating sufficient resources for \ac{ML} inference to the \acp{RIC} is going to be critical as they will have stringent delay requirements. 


\subsection{Intelligent Slice Orchestration and Management} 


The efficiency of the proposed semantic slicing framework will depend on intelligent slice orchestration and management. In the following, we discuss some of the challenges associated with the same.

\subsubsection{Continuous and distributed orchestration}

As user, application, and system domain contexts change with time, so must the semantic slicing orchestrator, requiring in-place continuous monitoring, collection of services states and
resources, \ac{AI}-driven analysis, \ac{CLA}, and service chaining. The proposed approach will utilize \ac{AI} models and semantics outputs to initiate and manage \ac{UE},
\ac{RAN}, edge, and core slices with required functional components (\ac{NSSF}, \ac{AMF}, \ac{UPF} etc.). 


\subsubsection{Heterogeneous service orchestration}

6G networks aim to support the convergence of physical, human,
and digital worlds within optimized and enhanced user’s \ac{QoE}, resulting in new and heterogeneous services with varying performance requirements. Semantic slicing will capture such performance requirements and provide additional context decomposition to describe relations between services and interfaces.

\subsubsection{Multi-stakeholder and multi-tenant orchestration}

Networks are expected to support the service continuum across multi-stakeholders and multi-tenant setups. The proposed \ac{E2E} semantic slicing orchestration will provide
interaction interfaces, switching mechanisms, and workflows, and will provision functional elements dynamically to support automated and secure coordination between stakeholders’ capabilities and resources. To this end, \ac{ORAN} can aid in the sharing of data across organizations through open interfaces to build a combined semantic knowledge base which can enable enhanced slice management over shared resources.

\subsubsection{Advanced monitoring and situational awareness} advanced, dynamic, and continuous monitoring will be required at the different
layers and interfaces of the network to enable \ac{CLA} and slice initiation and management, performance metrics generation, proactive orchestration, real-time fault and anomaly detection,
alerting, general diagnostics, and active response when/where needed. Semantics can help in building a cross-user, cross-application, and cross-domain situational awareness that can aid the \ac{NSP} in the efficient utilization and distribution of slice resources. 


\subsubsection{Adequate resource allocation for default slices} Semantic-based identification of different classes of computing tasks and related \ac{QoS} metrics will require an initial default slice with dedicated computational capabilities. Resources allocation and optimization to identify what is sufficient for such a slice is an open research question.

\subsection{Inter-Slice Interaction}

Inter-slice interactions like switching, handover or mobility management of \acp{UE} between different slices is still an open research problem. Inter-slice switching refers to \acp{UE} changing slice without maintaining session continuity. Whereas, inter-slice handover or inter-slice mobility management is one step ahead, i.e., \acp{UE} seamlessly moving from one slice to another without interrupting its current session. Currently \ac{3GPP} does not support inter-slice handover. An efficient inter-slice handover would be possible in two steps: (i)  prediction of the requirement of slice-switching beforehand (based on the semantics) and (ii) handover to a new slice seamlessly ensuring session continuity. In the case of a network-triggered inter-slice handover, the prediction can be done using a separate network function at the \ac{CN}.
For smooth inter-slice handover, it is crucial to predict the user requirements beforehand. 
System optimization will also require the identification of common and shared tasks and data in the infrastructure. Hence, creating the more advanced orchestration of slices and interactions between slices. 

\subsection{Security Architecture}  

it is vital to adopt a robust and secure architecture for 6G networks due to the expanded attack surface, resulting from higher degrees of complexity, multiple points of demarcation, and supported use cases. \ac{E2E} semantic slicing will utilize dynamic embedding of semi-autonomous agents (e.g. \ac{VNF} at network layers to enable local and global coordinated security monitoring, detection, and response.
\subsubsection{Slice isolation}
Dependencies in 6G networks are inevitable due to the virtualized nature of the architecture, the many integrative features, and the service-oriented model. However, dependencies can introduce opportunities for failures in one layer or component (e.g. slices) to propagate and affect other dependent resources and services, which in turn can compromise other slices \& services. These failures can be caused by operational issues, security compromises, or anomalies. Proper policies and controls, such as slice isolation and monitoring, can help manage and mitigate many of these risks. 

\ac{NS} isolation was identified by \ac{3GPP} and \ac{GSMA} as a key capability to ensure the possibility of concurrent execution of multiple \ac{NSI}, without negative impacts to each other's \acp{KPI} or other network elements~\cite{gsm2021e2e}. Furthermore, when \ac{E2E} \ac{NS} are deployed across multiple domains and multi-operators, high levels of \ac{NS} isolation may be required at system layers such as management, control, and resources ~\cite{gsm2021e2e}. \ac{NS} isolation can be categorized into three different types: (i) performance, (ii) management, and (iii) security/privacy isolation. Standardization bodies like \ac{GSMA} have already started to identify the need for slice templates to include the level of isolation as a parameter.

\subsubsection{Monitoring}
Trade-offs must be balanced between required global and local network visibility and controllability. Such trade-offs must be investigated when designing isolation mechanisms, policies, service chains, and enforcement points, especially as we adopt AI data-driven approaches for isolation that consider semantics. This must be supported by strong situational awareness at the different layers of the network, specifically for \ac{E2E} semantic slicing framework, which requires continuous and advanced monitoring. Basic capabilities of network monitoring functional components include: (i) collection, aggregation, and storage of real-time data generated by the different connected devices across the network continuum (at the \ac{CN}, edge/\ac{RAN} and \acp{UE}); (ii) performing real-time analysis; (iii) deriving insights and corresponding decisions in an autonomous and explainable manner; and (iv) execute in a decentralized and federated manner to support network management and service orchestration needs.

\section{Conclusion} \label{sec:concl}
Network slicing has emerged as one of the key enabling technologies to support new-age applications with diverse \ac{QoS} for \ac{6G} and next-generation wireless networks. However, the existing mechanisms for realizing network slicing are not able to bring out its potential completely due to some inherent limitations. This article presents the vision and a blueprint of a futuristic semantic slicing framework that goes beyond the boundaries of individual user, context, or application through cross-domain semantic knowledge sharing. It also paves the way for real-world deployment of semantic communication, which has recently been proven to enhance the traditional networks by crossing the Shannon limit, in cellular networks. We motivated this through some potential scenarios that might benefit from semantic slicing, provided a blueprint of the proposed framework, and also supported with a real-life application. Finally, we expand on interesting open research and implementation challenges for realizing semantic communication on 6G networks through semantic slicing. As part of future works, we would like to build on top of the initial results we have with the single-link baseline scenario and extend it to a more realistic multi-link use case and also strive for a proof-of-concept implementation of the proposed semantic slicing framework to realize the vision in its true sense.


\section*{Acknowledgment}
This work received support from the Commonwealth Cyber Initiative. For more information, see: \url{www.cyberinitiative.org}.









\bibliographystyle{IEEEtran}
\bibliography{IEEEtran/References}
\end{document}